\documentclass{article}
\usepackage{spconf,amsmath,graphicx}
\usepackage{caption}
\usepackage[]{algorithm2e}
\usepackage{listings}

\title{Computing on Masked Data to improve the Security of Big Data}
\name{Vijay Gadepally, B. Hancock and B. Kaiser,
  J. Kepner, P. Michaleas, M. Varia, A. Yerukhimovich\ \thanks{This work is sponsored by the
Assistant Secretary of Defense for Research and Engineering under
Air Force Contract \#FA8721-05-C-0002.  Opinions, interpretations,
recommendations and conclusions are those of the authors and are not
necessarily endorsed by the United States Government}}
\address{MIT Lincoln Laboratory, Lexington, MA 02420 \\ 
\{vijayg, braden.hancock, benjamin.kaiser, kepner, pmichaleas,
mayank.varia, arkady\}@ ll.mit.edu}

\begin{document}
%
\maketitle
%


\begin{abstract}
Organizations that make use of large quantities of information require the ability to store and process data from
central locations so that the product can be shared or distributed across a
heterogeneous group of users. However, recent events underscore the need for improving the security of data
stored in such untrusted servers or databases. Advances in cryptographic
techniques and database technologies provide the necessary security
functionality but rely on a computational model in which the cloud is
used solely for storage and retrieval. Much of big data computation
and analytics make use of signal processing fundamentals for computation. As the trend of
moving data storage and computation to the cloud increases, homeland
security missions should understand the impact of security on key signal
processing kernels such as correlation or thresholding.  In this
article, we propose a tool called Computing on Masked Data (CMD),
which combines advances in database technologies and
cryptographic tools to provide a low overhead mechanism
to offload certain mathematical operations securely to the cloud. This article describes the
design and development of the CMD tool.

\end{abstract}
\section{Introduction}
\label{sec:intro}

Recent events, such as~\cite{security}, highlight the growing need to improve the
security of big data stored in the cloud. The traditional challenges
associated with big data are often referred to as the 3 V’s of big data:
Volume, Velocity and Variety. The sensitivity of data
being processed in the cloud is increasing, and it is possible to make
use of cryptographic advances to secure processing in the cloud to
address a growing 4th V of big data - Veracity (the closest word to
security that starts with a 'V'). Unfortunately, the volume, velocity and variety properties of
big data systems challenge various system components and are often used to
justify ignoring this increasingly important factor. In a common
computing infrastructure where sensor data is fed into a shared
computing cloud to be used by authorized users, the goal is to
determine techniques that allow the potentially
untrusted computing cloud to store and perform computation directly
on encrypted or masked data.

Large heterogenous organizations, such as the Department of Homeland Security, have a particular need for making use of
the cloud. Key features such as resource pooling, rapid elasticity and
on-demand self-service~\cite{nist} provide the ability for diverse
organizations to share information efficiently and easily. However,
these advantages also create a single point where adversaries may try
to attack the confidentiality, integrity and/or availability of
sensitive information stored or being processed. As a specific
example, a recent report by the Department of Homeland Security (DHS)
looked at the viability of making use of cybersecurity insurance
(http://www.dhs.gov/publication/cybersecurity-insurance). One of the
requirements for evaluating the applicability of such a system will be in the ability to track cyber
incident reports in a central location such as a commercial cloud
offering. This data stored in the repository may provide
adversaries with information that could be used to reverse engineer
the details of a cyber security breach~\cite{nistcyber}. In this example and many more,
cryptographic protections can be used to mitigate the risks
of data being stored and processed in the cloud.

Prior work has looked at various techniques to improve the
confidentiality, integrity and availability of big data
systems and an overview is provided in~\cite{survey}. There are many existing cryptographic tools capable
of performing some level of computation on encrypted data. One of
these is Fully Homomorphic Encryption (FHE), first introduced in
2009~\cite{gentry2009fully}. This method allows arbitrary analytic
computations to be performed on encrypted data while maintaining a strong security
guarantee called ``semantic security'' (also referred to as randomized
encryption and is denoted as RND). The current
state of the art implementations of this scheme~\cite{halevi2014algorithms,perl2011poster} have a minimum
computational overhead of $~10^{5}$, which removes them from consideration
for any practical big data system. Secure Multi-Party Computation
(MPC) techniques~\cite{ben1988completeness,yao1982protocols} are another possibility with semantic security
guarantees, but these techniques also currently experience excessive
overhead and require large amounts of customization for a given
application and threat model. 



An alternative solution is to make use of more relaxed forms of
encryption that reveal small amounts of information, as first
described for the database setting in~\cite{popa2011cryptdb}. In such a system, one may make use of the deterministic
encryption method (DET), which encrypts any
plaintext and key to a single
ciphertext thereby allowing for fast equality
calculations to be performed on encrypted data (a property very
desirable for database applications). Other relaxed forms of
encryption, such as Order Preserving Encryption (OPE) which preserves
the relative order of all inputs,
provide high functionality in return for relaxed security guarantees. In this article we propose the
Computing on Masked Data (CMD)~\cite{kepner2014computing} system that
uses a combination of RND, DET, and OPE modes to mask sensitive data,
insert into a NoSQL database and perform simple computation on the
masked data. Further, for many situations, data integrity (not
necessarily confidentiality) is a large
concern and we provide a 4th masking level  - Authentication (AUT).
The article is structured as follows. Section~\ref{sec:cmd} provides
an overview of the proposed tool. Section~\ref{sec:implementation}
describes the implementation of the CMD
tool. Section~\ref{sec:sigproc} describes how CMD can be applied to
simple signal processing kernels which are important to analytic
developers. Section~\ref{sec:perf} discusses the performance of CMD on
these signal processing kernels. Finally, in Section~\ref{sec:conc} we
discuss potential CMD applications and
conclude.

\section{Computing on Masked Data}
\label{sec:cmd}

\begin{figure}[t!]
\centerline{
\includegraphics[width=3.5in]{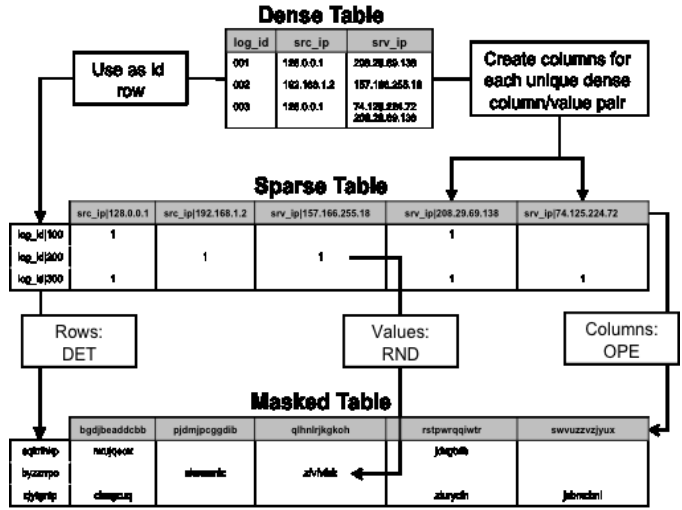}
}
\caption{Masking illustrative network data records.  Dense data is
  made sparse using the D4M schema and moves semantic information to
  the rows and columns of sparse assosciative arrays.  The sparse
  table is then masked using a variety of encryption schemes depending
  on the desired functionality. The terms DET, OPE and RND are
  described in Section~\ref{sec:implementation}.}
\label{cmd}
\end{figure}

Associative Arrays, the data type provided by the Dynamic Distributed
Dimensional Data Model (D4M)~\cite{kepner2013d4m}, are used as the base schema to represent
unstructured large data sets. The D4M schema encapsulates all semantic
information of a dataset into the rows and columns of a sparse
associative array. We make use of this property to use various different
encryption techniques to mask the rows and columns depending on
required functionality. An overview of the schema is described in
Figure~\ref{cmd}. 

Associative array rows, columns and values are masked through the CMD
mask function, which encrypts entities of the sparse associative
array. For example, if one intends to perform range operations on
columns, columns may be masked using OPE. Associative arrays support
the full range of linear algebra operations such as addition,
multiplication, etc. A more thorough description of the CMD system is
provided in~\cite{kepner2014computing}.

\section{CMD Implementation}
\label{sec:implementation}

A prototype of CMD has been developed as a MATLAB/GNU Octave
toolbox. The current version of CMD allows a user to select among
four masking levels: RND, DET, OPE and AUT. A combination of masking
levels can be used across the rows, columns and values of the
associative array representation. Selecting a masking level
depends upon the desired functionality. A summary of database (DB)
functionality and security afforded for different masking levels is provided in
Table~\ref{maskinglevel}.  The implementation of each of the
encryption techniques is described below.


\begin{table}[t!]
\centering
\begin{tabular}{|p{1.5cm} | p{3.5cm} | p{2.5cm}|}
\hline Masking Level & Information Leaked & Database  \mbox{     } Functionality\\ \hline 
RND & Hides all information about data & Decrypt only \\ \hline
DET & Reveals equality patterns & Query plaintext\\ \hline
OPE & Reveals equality patterns and order & Plaintext range query \\ \hline
AUT &Reveals everything but protects data
integrity & All queries \\ \hline
\end{tabular}
\caption{Summary of security provided and database functionality for
  CMD masking levels}
\label{maskinglevel}
\end{table}

\subsection{Semantically Secure Encryption - RND}

\begin{figure}[t!]
\centerline{
\includegraphics[width=3.5in]{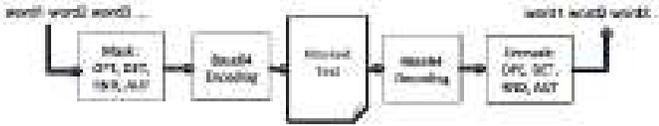}
}
\caption{Plaintext data is masked using a combination of possible
  modes. Masked data is then encoded for storage purposes. Accessing
  the data involves first decoding the stored masktext and unmasking.}
\label{process}
\end{figure}

The CMD implementation of RND uses the AES-256 block cipher found
in OpenSSL (www.openssl.org) in the Cipher Block Chaining or
Galois Counter Mode. The cryptographic key is derived
  from a user-provided password and an 8-byte salt using 1000
  rounds of the derived key generation loop. Each ciphertext has a minimum length of one AES block and is derived from both the cryptographic key and an
  Initialization Vector (IV) of 16 bytes. The IV is generated using
  the OpenSSL command \texttt{RAND\_bytes}, which generates an arbitrary length
  string of cryptographically strong pseudo-random bytes. This step ensures
  that each masking of the same plaintext will yield a different
  ciphertext, thereby protecting equality information. To simplify
  string handling, ciphertexts are converted into printable characters
  using Base64 encoding. An overview of the process is described in Figure~\ref{process}.

\subsection{Deterministic Encryption - DET}

The implementation of DET is identical to that of RND except for the
way that the IV is generated for each message. To leak only equality
information and nothing more, DET masking requires that each message
uses an IV that is unique to that message and deterministically
obtained. This is achieved in CMD by using the SHA-1 hash of each
message as its own IV, truncated to 16 bytes. The OpenSSL SHA-256
implementation may be substituted for SHA-1 if more security is
desired, but this comes with an approximate 40\% increase in
computation time compared to using the SHA-1 hash (which can perform
approximately 1.7x$10^{6}$ hashes/second).

\subsection{Order Preserving Encryption - OPE}

\begin{figure*}[ht!]
\centerline{
\includegraphics[width=7in]{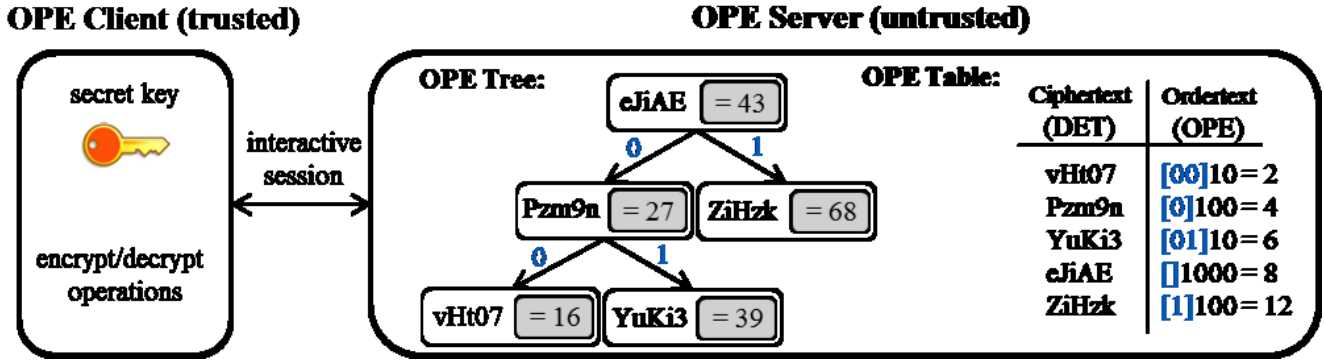}
}
\caption{Overview of the data structures in the mOPE scheme. Each node
  in the OPE tree contains a ciphertext. The plaintext value of each
  ciphertext is shown here in gray blocks for illustration purposes
  and are not stored in the tree. The ordertext value corresponding to
  each ciphertext comes from following the path to each ciphertext
  from the root of the tree. Ciphertext-ordertext pairs are stored in
  a database to enable fast lookup. Figure concept from~\cite{popa2013ideal}.}
\label{ope}
\end{figure*}

For the OPE masking level, a mutable scheme called mOPE, initially
proposed in ~\cite{popa2013ideal}, was adopted. The mOPE model consists of a trusted client and an untrusted server
that interact with each other. The untrusted server is never given
access to any plaintext values or the user password, and the trusted
client is never required to store or analyze the entire data set at
once.  The data are stored on the server in a binary search tree as
ciphertexts (obtained through DET encryption operations on the
client). Because ciphertexts do not leak order information, the server
must communicate with the client through an interactive session to
determine the correct location in the binary search tree for each
ciphertext. Starting at the root of the tree, the server sends the
client the ciphertext of the current node. The client decrypts the
ciphertext at that node, compares it to the plaintext being inserted,
and returns a ‘0’ (left) if the plaintext is less than the value at
that node or a ‘1’ (right) if the plaintext is greater. The new
current node is then returned, and the process repeats. When the exact
location of the plaintext to be inserted has been found, the client
sends the server the ciphertext of that value along with a command to
insert it there in the tree. The OPE ordertext representation of a
given plaintext is simply the path to its ciphertext in the binary
search tree, concatenated with padding of a ‘1’ and a sufficient
number of ‘0’s to make all ordertexts the same size. We once again
chose to use a default size of 16 bytes (thereby allowing for 216
entries if ‘0’s and ‘1’s are stored as strings or 2128 entries if the
path is stored as bits). Figure~\ref{ope} describes a sample OPE Tree and
corresponding OPE Table, which indicates the ordertext value for each
ciphertext in the tree. Querying for data (or a range of data) occurs
by determining the position of the masked data.

\subsection{Authentication mode - AUT}

The fourth supported operation available in CMD is AUT which stands for
“Authentication”. This operation does not actually mask the data at
all and plaintext is left as plaintext. Prepended to the plaintext,
however, is a hash-based message authentication code (HMAC). Here too,
we prefer to use SHA-1
for performance reasons. When a given message is unmasked with AUT,
the HMAC that is stored with the plaintext can be extracted and
compared against a new HMAC of the plaintext message and key. If the two are equal, the user
can have confidence that the original message has not been modified
since it was first stored as generating the same HMAC requires
knowledge of the key. While the AUT mode does not ensure data
confidentiality, it does ensure data integrity.

\section{Signal Processing with CMD}
\label{sec:sigproc}

In this section, we demonstrate the applicability of CMD on some basic
signal processing kernels. Given the relationship between the
associative array representation and graphs, we pay particular
attention to correlation and thresholding, which can be used as
building blocks to construct a variety of linear algebra and signal
processing algorithms.

\begin{figure}[h]
\centerline{
\includegraphics[width=3.45in]{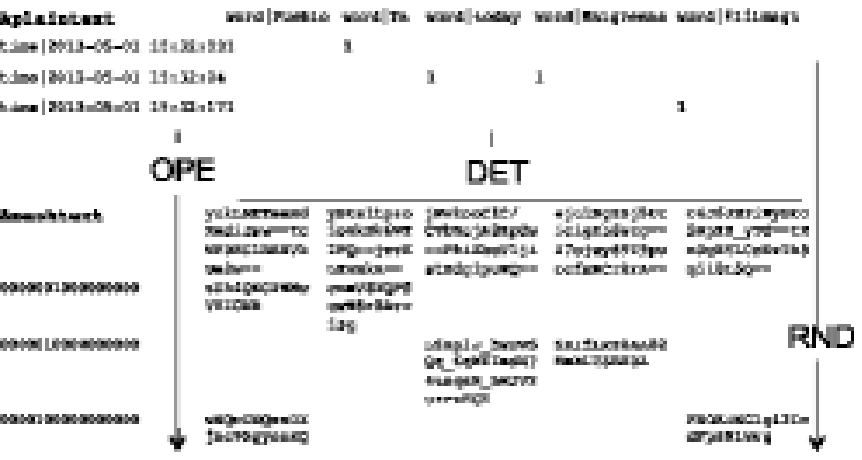}
}
\label{Twit_amt}
\end{figure}

Correlation provides a measure of the statistical relationship between
two entities. Calculating the correlation of two associative arrays
involves transposing and multiplying. Since CMD preserves the
associative array structure of big data, we can perform masked
correlation by multiplying two masked arrays which are masked using
any of the supported masking modes. This task can be done on a
potentially untrusted server. Consider a sample dataset which contains data collected from
the social media website Twitter. The data is stored as a tuple that contains a
timestamp, unique Twitter identification number (TweetID), and
value. A common task is to find the most common words that exist
within the same tweet as a term of interest, say \texttt{t1}. 

CMD can determine the unique TweetIDs by multiplying the masked
version of the dataset \texttt{Amasktext} (abbreviated as \texttt{Amt}) to yield an
output associative array \texttt{Cmasktext} (abbreviated as \texttt{Cmt}):

\begin{verbatim}
Cmt = Amt(:,Mask(t1))' * Amt
\end{verbatim}

Suppose that the term of interest is the word \texttt{happy}, which is
stored using the D4M schema as \texttt{word|happy}. The above product on
the \texttt{Amt} yields the following result:

\begin{figure}[h!]
\centerline{
\includegraphics[width=3.45in]{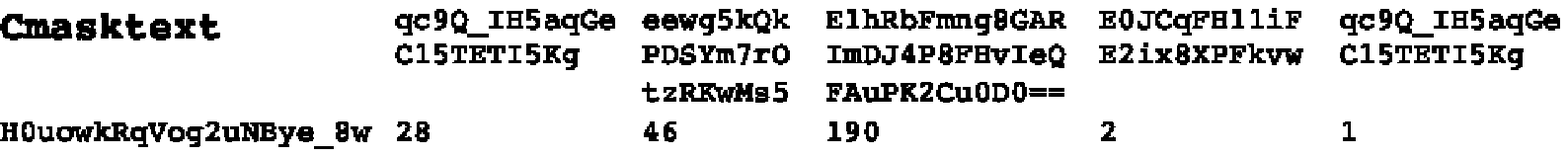}
}
\end{figure}

Unmasking \texttt{Cmasktext} reveals words correlated with \texttt{word|happy}:

\begin{verbatim}
Cpt=Unmask(Cmt,'DET','DET','RND')
\end{verbatim}

\begin{figure}[h]
\centerline{
\includegraphics[width=3.45in]{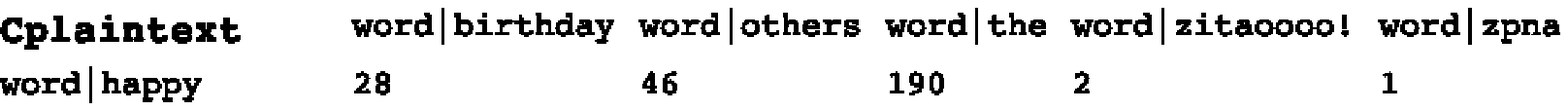}
}
\end{figure}

The second signal processing kernel currently supported is
thresholding. Thresholding keeps all values that satisfy a certain
criteria and discards the rest. In the example above, we may perform a
thresholding to find only common values. For example:

\begin{verbatim}
Cpt=Unmask(Cmt>20,'DET','DET','RND')
\end{verbatim}

\begin{figure}[h]
\centerline{
\includegraphics[width=3.45in]{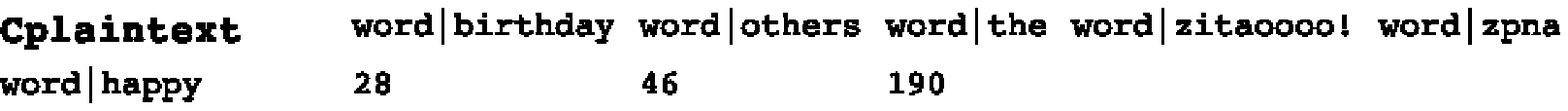}
}
\end{figure}

\section{Performance Results}
\label{sec:perf}

\begin{figure}[t]
\centerline{
\includegraphics[width=4in]{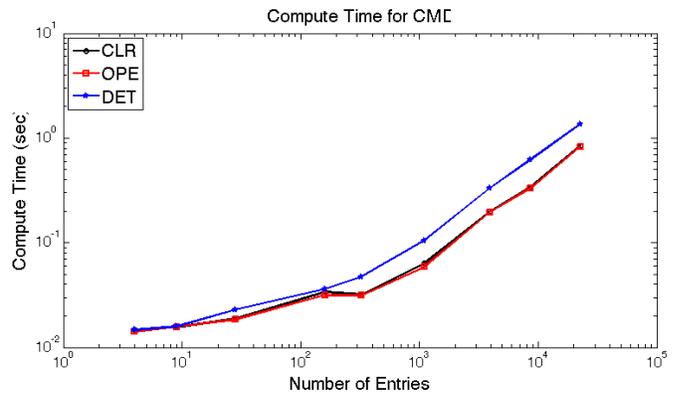}
}
\caption{Performance of different masking levels in performing a
  correlation of two arrays with varying number of elements}
\label{corrperf}
\end{figure}

This section describes the performance of the CMD system on varying
sizes of input data to quantify the relative performance overhead
incurred by performing computation on masked
data. For these examples, data is stored in an Apache Accumulo
NoSQL database and computation and data retrieval times are
presented. One key metric for CMD is the performance overhead incurred
with computation on masked data. Figure~\ref{corrperf} describes the time taken to compute a
correlation of two associative arrays with varying number of
entities. As the performance chart shows, the computational overhead involved with
using either DET or OPE modes of encryption is minimal (within 2x) when compared
to performing computation on the plaintext data (denoted by
CLR). Performing thresholding on masked data incurs almost no overhead.

\begin{figure}[h!]
\centerline{
\includegraphics[width=4in]{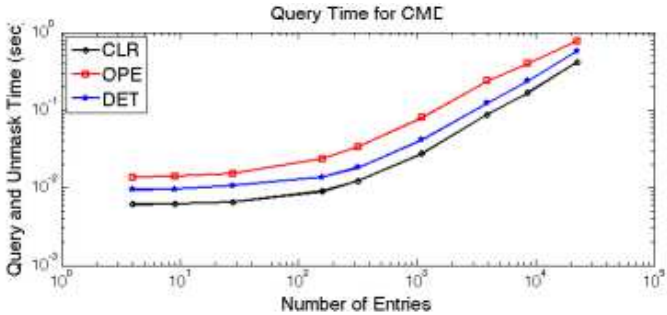}
}
\caption{Query and unmask performance time with varying number of entries}
\label{queryperf}
\end{figure}

Figure~\ref{queryperf} describes the performance overhead incurred in querying and
unmasking data masked using different levels of encryption. From this
figure, it is clear that the overhead incurred by using different
modes of encryption is similar to that of correlation (within 2x) when
compared to simply retrieving plaintext data (denoted by CLR).

While it is difficult to quantify the reduction in information leakage
when compared to plaintext data, the description of leakage in
Table~\ref{maskinglevel} shows that the CMD system greatly reduces
leakage for applications in which leakage of equality patterns or
order are acceptable.

\section{Discussion, Conclusions and Future Work}
\label{sec:conc}

Data veracity is a growing concern amongst the big data
community. The
volume, velocity, and variety, however, pose large challenges to the
development of systems capable of protecting the privacy of
big data. These challenges are especially evident in organizations
that operate large computing clouds to be shared by heterogeneous
organizations often with varying data policies. The Computing on Masked Data tool
presented in this article can be used in such environments. One
organization that exemplifies such an environment is the Department of
Homeland Security which is especially prone to such
challenges due to their distributed, multi-agency, and time sensitive
nature.  

Homeland security agencies are increasingly looking to cloud solutions
in order to make use of key
characteristics such as on-demand self-service, broad-network access,
resource pooling, rapid elasticity, and measured
service~\cite{nist}. Commercial entitites such as Amazon have
responded to this need by providing US based solutions such as the
Amazon GovCloud~\cite{govcloud}.

The Computing on Masked Data tool described in this article can
provide an extra layer of security with minimal impact to performance to individuals or organizations who
wish to make use of cloud computing offerings without completely trusting the
provider. Consider an example in which different agencies of the DHS wish
to maintain a database of security related data such as agent field reports. In order to ensure that this data can be accessed by
different agencies, perhaps geographically distributed across the
United States, the DHS may wish to make use of a commercial cloud
provider such as Amazon Web Service (AWS). While offerings such as
GovCloud provide a baseline of security, one may not want to
offload data confidentiality protection to a third party via service
level agreements and instead make use of cryptographic protections on
the data (thus protecting it from unauthorized users from the
untrusted cloud provider or other cloud tenants). Using CMD, generated
field reports can be encrypted at individual clients and uploaded to
the commercial cloud in encrypted form using the Mask command. Simple
computation such as correlation or thresholding (suppose to correlate
a name across reports from different agencies) can be done directly in
the cloud without compromising the confidentiality of requested
data. Finally, correlated data can be queried and returned to an
authorized end user who can use the Unmask command on the data using a valid key to
see the end result. Of course, there are still challenges associated
with such tools. For example, organizations may need to agree upon the
desired functionality in order to use the same choice of masking
levels. Such a challenge may be overcome using a layered encryption
scheme as described in~\cite{popa2011cryptdb}. Another challenge is
with the efficient management of keys and requires a solution as
proposed in~\cite{khazan2012lincoln}.

Homeland security researchers and scientists should be aware of the
increasing value of data in the cloud and determine methods to
mitigate the risk associated with unauthorized disclosure. In this article,
we propose a tool - Computing on Masked Data - that is a high
performance NoSQL database interface that stores data masked by a
choice of techniques. The CMD system supports a wide variety of
masking modes and exhibits high performance for common signal processing kernels such as
correlation and thresholding. The performance and simplicity of the
CMD tool make it an ideal candidate for many
applications.

\section{Acknowledgements}

The authors would like to acknowledge the anonymous reviewers, Nabil Schear, Rob Cunningham, and the LLGrid operations team at MIT Lincoln
 Laboratory for their support in developing and testing the CMD prototype.

\vfill

\bibliographystyle{IEEEbib}
\bibliography{ieeehst}

\end{document}